\documentclass[preprint,12pt]{elsarticle}

\usepackage{amssymb}

\usepackage{siunitx}
\usepackage{caption}
\usepackage{subcaption}
\usepackage{algpseudocode}
\usepackage{algorithm}
\usepackage[table,xcdraw]{xcolor}
\usepackage{fancyvrb}
\usepackage{hyperref}

\usepackage{lineno}

\journal{Nuclear Physics B}

\begin{document}

\begin{frontmatter}



\title{Accelerated ray-tracing simulations using McXtrace}


\author[DTUP,Exruptive]{Steffen Sloth}
\author[DTUP,ESS]{Peter Kjær Willendrup}
\author[DTUC]{Hans Henrik Brandenborg Sørensen}
\author[Exruptive]{Morten Christensen}
\author[DTUP]{Henning Friis Poulsen}

\affiliation[DTUP]{organization={Department of Physics, Technical University of Denmark},
            addressline={Fysikvej 1}, 
            city={Kongens Lyngby},
            postcode={2800}, 
            state={Capital region},
            country={Denmark}}

\affiliation[Exruptive]{organization={Exruptive A/S},
            addressline={Højnæsvej 75}, 
            city={Rødovre},
            postcode={2610}, 
            state={Capital region},
            country={Denmark}}

\affiliation[ESS]{organization={ESS DMSC},
            addressline={Asmussens Allé 305}, 
            city={Kongens Lyngby},
            postcode={2800}, 
            state={Capital region},
            country={Denmark}}

\affiliation[DTUC]{organization={Department of Applied Mathematics and Computer Science, Technical University of Denmark},
            addressline={Richard Petersens Plads 324}, 
            city={Kongens Lyngby},
            postcode={2800}, 
            state={Capital region},
            country={Denmark}}
            
\begin{abstract}
McXtrace is an established Monte Carlo based ray-tracing tool to simulate synchrotron beamlines and X-ray laboratory instruments. This work explains and demonstrates the new capability of GPU-accelerated McXtrace ray-tracing simulations. The openACC implementation is presented, followed by a demonstration of the achieved speed-up factor for several types of instruments across different types of hardware. The instruments achieve speed-up factors around \SIrange{250}{600}{} dependent on the instrument complexity. Instruments requiring repeated memory access might require optimised memory access procedures to avoid severe penalties in the simulation time when using GPUs. The importance of reducing the simulations was demonstrated for an aviation security application by comparing the simulation time of a projection of an energy-dispersive X-ray computed tomography instrument.

\end{abstract}




\end{frontmatter}


\section{Introduction}
\label{sec:1}

This work focuses on accelerating ray-tracing simulations. Investigations using X-rays have been implemented in a wide range of industries and research fields, such as medical science \cite{medsci_1,medsci_2}, aviation security \cite{avisec_1}, manufacturing \cite{manufac_1}, and material science \cite{matsci_1}. A successful X-ray experiment relies on an understanding of the X-ray interaction mechanisms between X-rays and matter together with an insight into the instrument's capabilities. Such insight can be challenging to achieve for modern X-ray techniques, due to the increased complexity of the measurement technique and instrument. Simulations can serve to predict the experimental outcome and assist researchers in designing and verifying results before commencing a costly experiment. Simulations can also assist in evaluating and enhancing the instrument’s performance and capabilities. \\
\\
The choice of simulation tool is important since their capabilities can vary greatly. Simulation tools can be designed for task-specific applications, such as radiation dose reduction in medical X-ray imaging \cite{medsci_1}, or to model general X-ray applications. Tools for general application can model the X-ray interactions at various levels of detail. Simple models like Single-Point Ray Tracing \cite{debisim} and X-ray forward modelling \cite{NoviSim} achieve short simulation times but at a reduced level of detail and simulated realism. Achieving a higher level of realism requires more detailed modelling of the X-rays and the interaction mechanisms, as done by forward ray-tracing and wavefront propagation software \cite{mcx_1, shadow}, but at the cost of increased simulation times. The computational time is a compromise between the available computing power and the problem's complexity. \\
\\
McXtrace, a forward ray-tracing simulation tool based on the Monte Carlo method, has been used since 2009 to design both instruments and experiments \cite{mcx_1}. A high level of realism can be achieved using the McXtrace framework. A McXtrace simulation is performed by compiling an instrument file, which in principle is a sequential list of the needed X-ray components. The properties and geometrical configuration of each X-ray component are described in the McXtrace instrument file. The McXtrace software comes with a wide range of predefined X-ray components, including X-ray sources, optics, samples, and detectors. The open-source nature of McXtrace also allows users to define their own custom components. \\
\\
A ray-tracing simulation requires a large number of rays to be generated and traced through the instrument. In McXtrace the rays are generated at the source component and sequentially propagated through the list of X-ray components. This process is slow since each ray’s interaction with each X-ray component needs to be computed, which for even simple instruments can lead to tens of millions of calculations. The improved modern computational hardware reduces the computation time and thereby allows for more complex and computation-heavy simulations to be performed. \\
\\
There still is an increasing desire for simulating more complex scenarios with even more rays for improved realism and simulation accuracy. Even the performance of modern CPUs struggles to keep up with the demand for computing power. An example of such a use case is the current development and implementation of energy-dispersive X-ray detectors, where the simulations get an additional spectral dimension which needs to be resolved \cite{mcx_2}. X-ray systems using Photon Counting Detectors (PCDs) to detect the X-ray’s energy are generally more challenging to design and manufacture \cite{edxct_1}. The currently available PCDs are limited in size and resolution, both spatially and spectrally. Modelling the capabilities of potential instrument designs for energy-dispersive X-ray systems using simulations is greatly desired to reduce cost and increase instrument performance. \\
\\
Simulations of energy-dispersive instruments are already possible in McXtrace models. However, since each ray is monochromatic, a larger number of rays is needed to probe the additional spectral dimension. Even utilising CPU parallelisation, the simulation times turn out to be in the range of days and weeks. Furthermore, the increased use of X-ray Computed Tomography (CT) instruments in industrial settings has sparked the need for simulation CT acquisitions, where many radiographies must be produced \cite{mcx_2}. New computational enhancements are needed to accommodate the requirements of simulating energy-dispersive and CT instruments using McXtrace. \\
\\
This paper presents a new development in McXtrace to utilise GPU parallelisation to enhance computational performance and thereby reduce simulation times. The McXtrace framework and the GPU implementation are described in section 2. Section 3 presents three types of simulation problems with increasing complexity together with a description of a McXtrace instrument for each of the problems. Section 4 presents the simulation time and relative speed-up factors for simulations reproduced using a list of different hardware configurations at an increasing level of count statistics. The required level of count statistics for performing energy-dispersive simulations and the estimated simulation times are discussed in section 5, together with an observed memory issue and how GPU-accelerated simulations are expected to generalise for other McXtrace instruments.

\section{The McXtrace framework and OpenACC GPU support}
\label{sec:2}

As mentioned in the introduction, McXtrace is a general X-ray instrument simulation framework written in C. McXtrace \cite{mcx_1}-\cite{mcx_2} and its older sibling McStas\cite{mcstas_1}-\cite{mcstas_3} both implements a Monte Carlo ray-tracing approach based on a domain-specific language (DSL) model, i.e. a dedicated programming language, implemented as a text-parser and a c-code generator using the Lex and Yacc tools\cite{ibm_lex_yacc}.

\subsection{Using McXtrace}

The user of the simulation tool writes a so-called \emph{instrument} file using the specific McXtrace grammar, thereby arranging the physical objects (implemented in so-called \emph{components}) as they are placed relative to each other in the simulated lab-space. \\
\\
The McXtrace library currently contains 205 components sorted in categories, 88 example instruments and an archive of datafiles:

\begin{itemize}
    \item[\texttt{sources}:] Contains components defining the initial X-ray beam at the source, including everything from a monochromatic, mathematical point-source over models of lab-sized X-ray sources to undulators and insertion-devices at large-scale X-ray facilities.
    \item[\texttt{optics}:] Contains X-ray optical elements, such as mirrors, lenses, capillaries, monochromators and zone-plates.
    \item[\texttt{samples}:] Contains models of \emph{matter} that may interact with the beam, e.g. for diffraction, spectroscopy and imaging - often with material definitions from tabular input, also included with McXtrace.
    \item[\texttt{sasmodels}:] This special category contains many models for small-angle scattering from \emph{matter}, derived from SasView\cite{sasview},\cite{mcstas_sasview}.
    \item[\texttt{monitors}:] Contains \emph{measurement} components, allows to measure properties of the beam at the given point in the instrument, e.g.: energy-spectrum, divergence, beam cross-section. Most of these models are not physical, i.e. include no measurement-related losses and no description of detector efficiency.
    \item[\texttt{union}:] Framework components to assemble arrangements of material-geometries, to allow describing e.g. a sample within a sample container within a sample-environment.
    \item[\texttt{astrox}:] Components for modelling X-ray telescopes.
    \item[\texttt{misc}:] Miscellaneous other components that are none of the above, e.g. for producing or reading particle lists in MCPL\cite{mcpl} format.
    \item[\texttt{contrib}:] Components written by users of McXtrace, i.e. not developed by the McXtrace team. 
    \item[\texttt{obsolete}:] Older components that are still useful but where newer components in the above categories are recommended. Often the obsoleted components are still in use within one or more of the McXtrace example instruments.
    \item[\texttt{examples}:] Contains McXtrace example instruments from a wide range of applications.
    \item[\texttt{data}:] Folder with datafiles for McXtrace, e.g. source parameter descriptions, material definitions etc.
\end{itemize}

During a McXtrace simulation, an ensemble (e.g. \SIrange{e6}{e9}{}) of rays are transported from an X-ray source, through an instrument or beam-line composed from the above selection of components, and at given points the user may set up to measure properties of the beam. Thus, as an example, the simplest imaginable McXtrace instrument would contain a \emph{source} component to emit X-ray particles and a \emph{monitor} component e.g. measuring the spectrum emitted by the source:

\begin{Verbatim}[frame=single]
DEFINE INSTRUMENT simple(Ekev=50, dEkev=1)

TRACE

COMPONENT Source = Source_flat(radius=0.01, 
                               E0=Ekev, dE=dEkev)
AT (0,0,0) ABSOLUTE

COMPONENT E_mon = E_monitor(nE=100, Emin=0, Emax=100)
AT (0,0,0.001) RELATIVE Source
   
END
\end{Verbatim}

McXtrace includes a graphical user-interface (\texttt{mxgui}), a command-line simulation tool (\texttt{mxrun}), a tool for visualising the instrument geometry (\texttt{mxdisplay}) and a tool for plotting \emph{monitor} output (\texttt{mxplot}).

\subsection{Parallelism of McXtrace simulations}

The chosen statistic of beamlets is treated in an intrinsically independent fashion and thus poses an \emph{embarrassingly parallel}\cite{parallel} problem, highly suitable for parallel computing. 

\subsubsection{McXtrace MPI simulations}

For many years MPI McXtrace (and McStas) have supported parallel simulations using MPI\cite{mpi}. The MPI implementation is simple: Instead of calculating the full beamlet statistic serially on one processor, each of $N$ processors is tasked with $\frac{1}{N}$ of the initial calculation. We apply a \emph{scatter - gather} approach, i.e. the problem is initialised and split between the worker processes that each simulate their particle statistic share sequentially, and after completion of all processes all results are added/concatenated together.

\subsubsection{McXtrace GPU simulations}

The major version 3 of McXtrace has support for parallelisation using GPUs\cite{gpu}. The implementation uses OpenACC\cite{openacc} pre-processor directives, implemented where needed in our components and inserted automatically into the generated c-code from the McXtrace code generator. The most mature implementation of OpenACC is available from Nvidia through the NVHPC package\cite{nvhpc} and its \texttt{nvc} compiler. The McXtrace (and McStas) implementation for GPUs uses special features of the \texttt{nvc} compiler. Therefore, the implementation is at the moment restricted to that compiler and thus Nvidia GPU cards. Since Nvidia has so far only released NVHPC for Linux, this is the only platform where McXtrace can be GPU accelerated. \\
\\
GPU parallel execution means parallel execution of multiple threads on the GPU hardware. A major undertaking during the port of McXtrace to the OpenACC GPU solution was to avoid race conditions, i.e. access of multiple threads to the same memory block. Effectively all components have been revisited to ensure that e.g. all calculation intermediates are now local and that the few needed global structures (e.g. monitor arrays) are locked for access by a single thread at a time. \\
\\
The main structural elements of the McXtrace 3.x GPU implementation are:

\begin{enumerate}
    \item A modernised code-generator, based on functions. (E.g. to allow component instances of the same type to call the same underlying function. In earlier versions of McXtrace, the code instead contained duplicate code masked by precompiler \texttt{\#define} pragmas.)
    \item Use of automatically generated structs to hold component instance information, components representing physical elements of the instrument.
    \item A reimplementation of the random number generator for the GPU 
    \item A new \texttt{\_particle} c-struct to hold physical and other parameters of the particle:
    \begin{itemize}
        \item Spatial coordinates $x,y,z$ and time $t$
        \item Wavevector components $k_x,k_y,k_z$
        \item X-ray phase $\phi$ and electrical field components $E_x,E_y,Ez$
        \item X-ray Monte Carlo weight $p$
        \item Various simulation-flow control parameters and state of the random number generator (needed independently pr. thread on GPU)
        \item The new \texttt{USERVARS}, see below
    \end{itemize}
    \item A new instrument- and component-level syntax called \texttt{USERVARS} to allow particle-dependent flags (e.g. to monitor the number of scatterings etc.) as part of the \texttt{\_particle} struct.
    \item Components needed to become \emph{thread safe}, i.e. variables depending on the particle state must be of local scope.
    \item Plus essential instances of \texttt{\#pragma acc}
    \begin{itemize}
        \item \texttt{\#pragmac acc declare create(var)} - requests that \texttt{var} becomes available on the GPU
        \item \texttt{\#pragmac acc update device(var)} - requests that \texttt{var} on the GPU is updated from its CPU counterpart
        \item \texttt{\#pragmac acc update host(var)} - requests that \texttt{var} for the CPU is updated from its GPU counterpart
        \item \texttt{\#pragmac acc atomic} - instructs the compiler to \emph{lock} operations in the following code block to a single GPU thread. (E.g. for updating arrays of the monitors.)
    \end{itemize}
\end{enumerate}

\subsubsection{Combined McXtrace MPI-GPU simulations}

Enabling multiple GPUs was trivial since the existing MPI implementation was simply used to split the initial calculation problems into sub-problems that could then run on multiple GPUs in parallel.

\section{Method testing the performance of GPU-accelerate McXtrace simulations}

To illustrate the performance of parallelism of McXtrace, three cases were investigated with McXtrace instruments of increasing levels of complexity:
\begin{enumerate}
    \item \label{prob1} A basic instrument with source and monitor. 
    \item \label{prob2} An instrument including source, compound refractive lens (CRL) and monitors. 
    \item \label{prob3} A flexible instrument able to take on various geometry configurations and model energy-dispersive X-ray-Matter interactions of custom 3D sample geometries for computed tomography applications
\end{enumerate} 

This section describes the instrument configuration used in the three cases together with the measurement procedure and hardware devices.

\subsection{A basic McXtrace instrument}

\begin{figure}[h!]
     \centering
     \includegraphics[width=.7\textwidth]{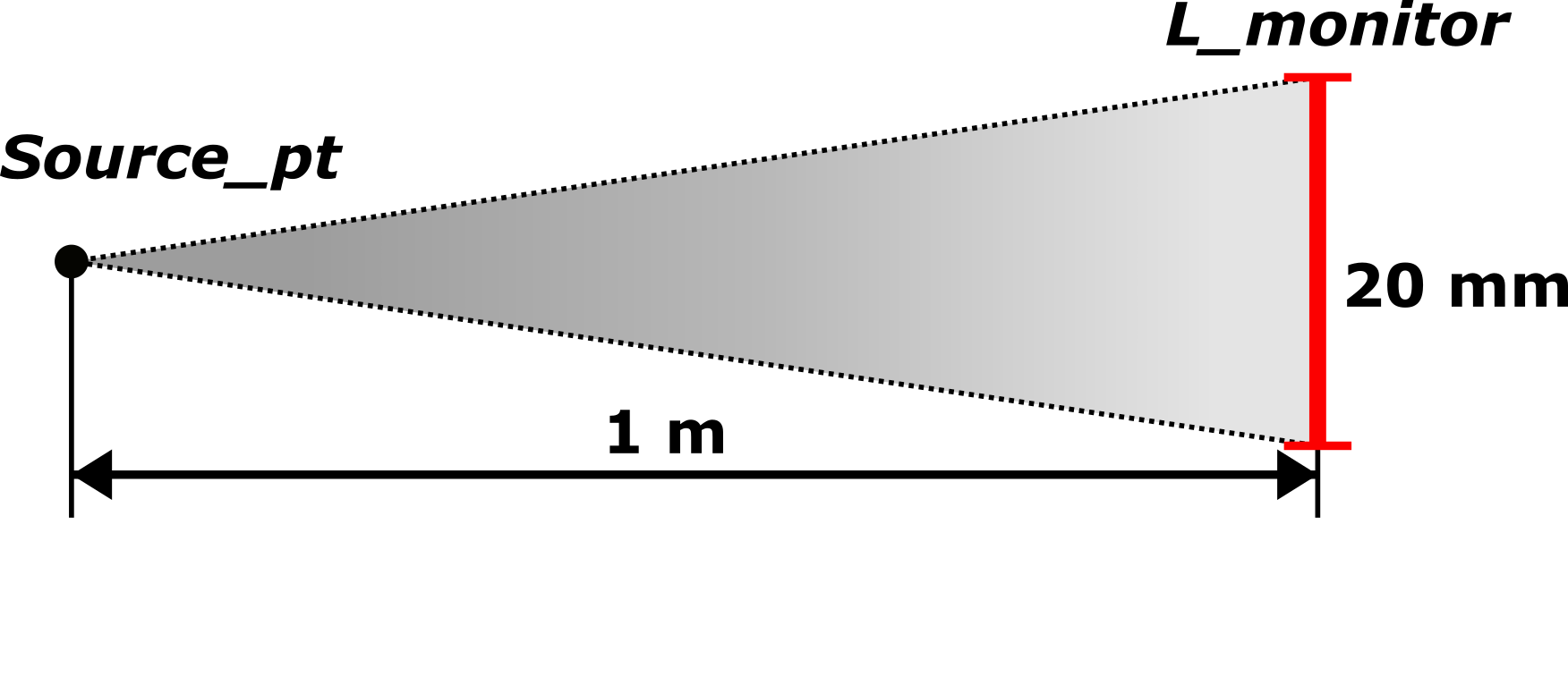}
     \caption{A sketch (not to scale) of the basic instrument. The cone beam emitted by the point source (\textit{Source\_pt}) was cropped to the size of the 2D $\SI{20}{\mm} \times \SI{20}{\mm}$ square \textit{L\_monitor}. The ray's wavelength was detected by the monitor.}
     \label{fig:3:1:basic_instr}
\end{figure}

The basic McXtrace instrument contains a source (\textit{Source\_pt}) and a monitor (\textit{L\_monitor}) component. The configuration of the instrument's geometry is shown in Fig.~\ref{fig:3:1:basic_instr} (not to scale). The \textit{Source\_pt} is a point source emitting a cone beam with a flat intensity profile centred at an X-ray energy of \SI{5}{\keV} and energy half spread of \SI{1}{\keV}. The source beam was cropped to a square window sized to illuminate the $\SI{20}{\mm}\times \SI{20}{\mm}$ monitor placed \SI{1}{\m} from the source. The \textit{L\_monitor} detects the X-ray wavelength, similar to how the energy-dispersive detector \textit{E\_monitor} detects the rays' energy.

\subsection{A focusing instrument using compound refractive lenses}

\begin{figure}[h!]
     \centering
     \includegraphics[width=.9\textwidth]{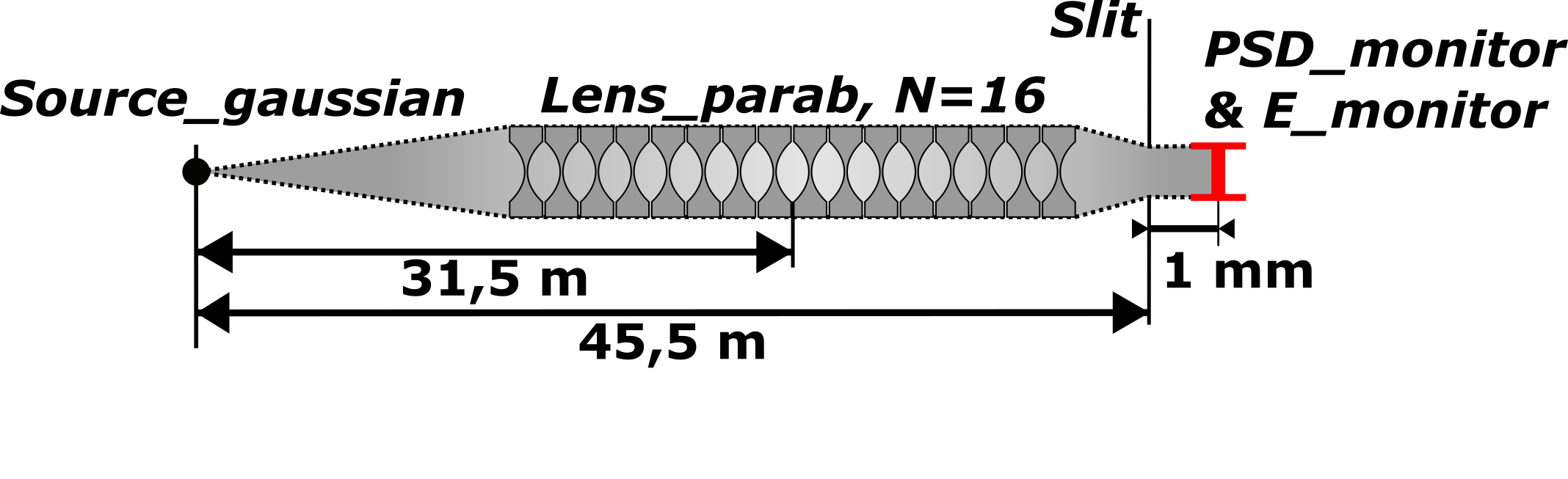}
     \caption{A sketch (not to scale) of the focusing instrument. The \textit{Source\_gaussian} models an undulator source, which beam was cropped to the size of the lens opening aperture. The lens was composed of a stack of sixteen double-sided parabolic beryllium lenses. Unfocused rays were absorbed (eliminated) by a slit placed \SI{14}{\m} after the lens centre. The ray's position and energy were finally detected using the two monitor components: \textit{PSD\_monitor} and \textit{E\_monitor}.}
     \label{fig:3:1:focus_instr}
\end{figure}

The components in the focusing instrument were aligned along the optical axis as shown in Fig.~\ref{fig:3:1:focus_instr}. The rays were traced from the source (\textit{Source\_gaussian}) through the lens component (\textit{Lens\_parab}), and finally the positions and energy of the rays passing the slit component (\textit{Slit}) were detector via two separate components (\textit{PSD\_monitor} and \textit{E\_monitor}). 

\begin{table}[h!]
\centering
\caption{The configuration parameters for the \textit{Source\_gaussian} component.} \label{tab:3.2:source}
\begin{tabular}{|l|c|} \hline
    \textbf{Component}            & \textit{\textbf{Source\_gaussian}} \\ \hline
    Horizontal standard deviation & \SI{48.2}{\um}      \\ \hline
    Vertical standard deviation   & \SI{9.5}{\um}       \\ \hline
    Horizontal divergence         & \SI{100}{\um}       \\ \hline
    Vertical divergence           & \SI{4.3}{\um}       \\ \hline
    Central energy ($E_0$)        & \SI{23.32}{\keV}    \\ \hline
    Energy half width ($\sigma$)  & \SI{1}{\keV}        \\ \hline
    \end{tabular}
\end{table}

The \textit{Source\_gaussian} component models a synchrotron undulator source emitting a Gaussian cone beam with its settings given in table~\ref{tab:3.2:source}. The cone beam was cropped to a $\SI{1}{\mm}\times \SI{1}{\mm}$ square window \SI{31.5}{\m} along the optical axis at the centre of the lens components. The CRL lens component \textit{Lens\_parab} models a stack of sixteen ($N=16$) parabolic lenses made of beryllium (Be) with a \SI{200}{\um} radius of curvature and lens-to-lens distance of \SI{50}{\um}. The \textit{Slit} component absorbs (eliminates) all the stray rays not being focused through its \SI{0.1}{\mm} circular opening aperture. Rays passing the slit have their position and energy detected by a Position Sensitive Detector (\textit{PSD\_monitor}) and an energy sensitive monitor (\textit{E\_monitor}). \\
\\
The majority of the simulation time was spent calculating the beam refraction of the CRL. The computation time spent on rays absorbed by the slit was wasted since these rays were eliminated.

\subsection{An energy-dispersive X-ray computed tomography instrument} \label{sec:3.3}

The simulated instrument was a digital clone of an Energy-Dispersive X-ray Computed Tomography (EDXCT) instrument for aviation security \cite{ss-euspen2023}. The development of Photon Counting Detectors (PCDs) has enabled the industrial implementation of energy-dispersive X-ray systems \cite{edxct_1}. PCDs are costly and the implementation often comes with a series of challenges. Therefore, simulating systems using PCDs is of great interest to evaluate and optimise the system's performance. The EDXCT instrument was designed to test the implementation of PCDs for enhanced material classification in aviation security. Accelerated simulation times are important when simulating EDXCT instruments since a large number of rays are required to probe both the energy and spatial domains.

\begin{figure}[h!]
     \centering
     \includegraphics[width=.9\textwidth]{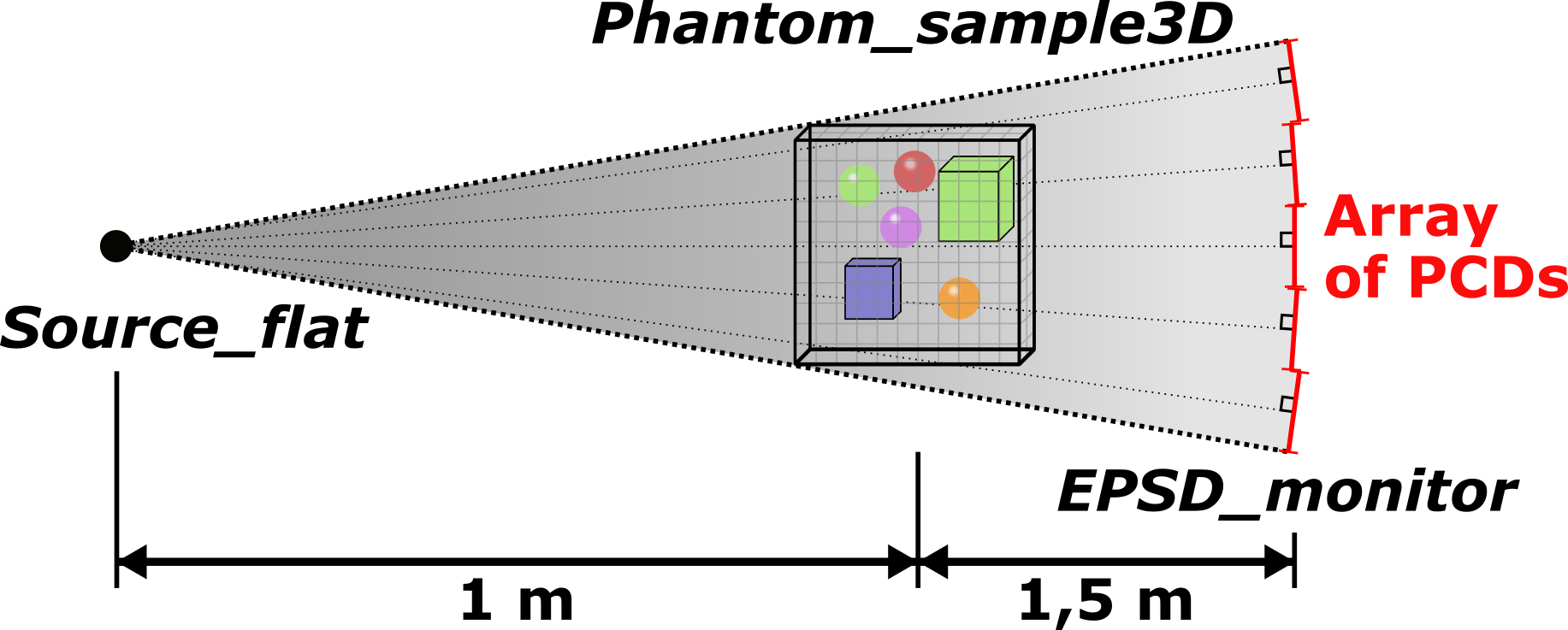}
     \caption{A sketch (not to scale) of the EDXCT instrument. A fan beam was emitted to illuminate the full array of five energy-dispersive detectors (\textit{EPSD\_monitor}). The \textit{EPSD\_monitor} components were oriented around \SI{2.5}{\m} from the source and tilted so a line from the source to the detector centre would be perpendicular to the detector's plane. The 3D  sample component was configured as a $\SI{30}{\cm}\times\SI{40}{\cm}\times\SI{55}{\cm}$ rectangle of water.}
     \label{fig:3:1:EDXCT_inst}
\end{figure}

The configuration of the EDXCT in-line instrument is shown in Fig.~\ref{fig:3:1:EDXCT_inst}. The X-ray source (\textit{source\_flat}) emits a polychromatic X-ray fan beam defined by the bremsstrahlung spectrum of a tungsten anode (in these simulations). The fan beam illuminates the full detector array, composed of five 1D (line) PCDs stacked vertically to cover a sufficient field of view. The PCDs were modelled by the Energy-Position-Sensitive-Detector (\textit{EPSD\_monitor}) component in McXtrace. The EPSDs were configured as the industrial PCDs X-Card ME3 manufactured by Detection Technology \cite{DTX}. The settings of the simulated source and detector components are given in table~\ref{tab:3:param}.

\begin{table}[h!]
    \centering
    \caption{The parameter configuration of the seven components (1 source, 1 sample, and 5 detectors) in the EDXCT instrument.} \label{tab:3:param}
    \begin{tabular}{|l|c|} \hline
        \textbf{Component}      & \textit{\textbf{Source\_flat}}                        \\ \hline
        Anode material          &  Tungsten                                             \\ \hline
        Focal spot size         &  $\SI{10}{\um} \times \SI{10}{\um}$                   \\ \hline
        Acceleration voltage    &  \SI{160}{kV}                                         \\ \hline
        \textbf{Component}      & \textit{\textbf{EPSD\_monitor} ($\times\SI{5}{}$)}    \\ \hline
        Pixel array size        & $\SI{1}{} \times \SI{128}{}$                          \\ \hline
        Pixel size              & $\SI{0.8}{\mm} \times \SI{0.8}{\mm}$                  \\ \hline
        Energy range            & \SIrange{20}{160}{\keV}                               \\ \hline
        \begin{tabular}[c]{@{}l@{}}Number of \\ energy bins\end{tabular} & \SI{128}{}   \\ \hline
        \textbf{Component}      & \textit{\textbf{Phantom\_sample3D}}                   \\ \hline
        Dimensions              & $\SI{350}{} \times \SI{250}{} \times \SI{100}{}$      \\ \hline
        Voxel size              & \SI{1}{\mm}                                           \\ \hline
        Geometry configuration  & Rectangle                                             \\ \hline
        Material                & Water                                                 \\ \hline
    \end{tabular}
\end{table}

The sample component \textit{Phamtom\_sample3D} \cite{mcx_2} was placed in between the source and detector array, as shown in Fig~\ref{fig:3:1:EDXCT_inst}. The component represents the sample as a 3D segmented volume, provided by the user, and models the energy-dispersive X-ray interaction mechanisms of absorption and scattering. A voxel in the 3D volume contains integer values (material indices) used to identify the relevant material properties from a \textit{struct} of every material in the volume. \\
\\
The sample component traces the rays in small steps through the segmented 3D voxel volume. Each step consists of three parts: 1) the ray was propagated forward by a small increment, 2) the material index was identified by finding the voxel containing the new ray positions, and 3) updating the ray's attenuation (Monte Carlo weight) and direction (wavevector) according to the relevant material properties. The size and resolution of the 3D volume were determined by the number of voxels in each dimension and the voxel size. The propagating step size was by default 1/3 of the voxel size. \\
\\
The majority of the simulation time was spent tracing the rays through 3D volume, since for luggage size volumes ($\SI{30}{\cm}\times\SI{40}{\cm}\times\SI{55}{\cm}$) with a millimetre voxel size the number of steps reaches into the thousands. \\
\\
Performing a CT simulation of this line scanner configuration requires a simulation for each imaged line (sample slice) repeated for each rotational position of the sample (projection). A \SI{55}{\cm} long sample with a \SI{1}{\mm} line width would require 550 lines per projection. The simulation time depends on the density of the non-empty voxel and the average probability of a scattering event (prolonging the ray's path length). The sample volume was a solid $\SI{30}{\cm}\times\SI{40}{\cm}\times\SI{55}{\cm}$ rectangle of water. This sample configuration maximises the number of ray interactions (computations) and results in an upper estimate of the simulation time.

\subsection{Hardware and procedure for estimating the simulation time and speed-up}

Testing and comparing the performance of McXtrace simulations were done by running simulations of the three instruments, described above, for an increasing number of rays and comparing the simulation time with simulations reproduced across different computational hardware. The simulation was run for several problem sizes with ray statistics between \SIrange{1}{e9}{}. A list of the computational hardware used for repeated simulations is given in table~\ref{tab:hardware}. Simulations using the CPU devices were run as single-threaded and multi-threaded processes using MPI.

\begin{table}[h!]
    \centering \small
    \caption{A list of the four hardware configurations used to test the simulation and compare the simulation times. The number of cores (column three) refers to the number of double-precision cores (FP64). The number of GPU threads scales linearly with the number of rays (\#Rays) up until \SI{2e9}{}.} \label{tab:hardware}
    \begin{tabular}{|l|l|c|c|c|c|}  \hline
    \textbf{Type} & \textbf{Name} & \textbf{Cores} & \textbf{LL Cache} & \textbf{Bandwidth} & \textbf{Threads} \\ \hline
    CPU  & \begin{tabular}[c]{@{}l@{}}Intel Xeon E5\\-2687W v4 \end{tabular} 
                                     & 12    & \SI{30}{MB}    & \SI{76.8}{GB/s}   & 1, 10   \\ \hline
    CPU  & \begin{tabular}[c]{@{}l@{}}AMD Ryzen Thread-\\ ripper PRO5975WXs \end{tabular} 
                                   & 32    & \SI{128}{MB}   & \SI{204.8}{GB/s}  & 1, 10, 32 \\ \hline
    GPU  & \begin{tabular}[c]{@{}l@{}}NVIDIA GeForce\\ GTX 1080 \end{tabular} 
                                   & 80    & \SI{2}{MB}     & \SI{320.0}{GB/s}  & \#Rays    \\ \hline
    GPU  & NVIDIA H100             & 7296  & \SI{50}{MB}    & \SI{2039.0}{GB/s} & \#Rays    \\ \hline
    \end{tabular}
\end{table}
 
The simulation time was measured via the machine wall clock so that the time estimates would be close to what regular users would experience. To illustrate the advantage of improved computational hardware (multi-core CPUs and GPUs) the simulation time was converted into a speed-up factor by dividing the measured times by a baseline. The baseline was defined as the simulation time for the Intel Xeon CPU workstation PC run as a single-threaded process. \\
\\
The time measurement variations were estimated by repeating the simulations $M$ times for each number of rays on each hardware setting. For the basic and focusing instrument $M=100$ was used for each simulation. The EDXCT had longer simulation times and the number of repetitions varied between hardware and the number of rays to keep the total simulation time below \SI{e4}{\s} (\SI{24}{\hour}). For the EDXCT simulation, the time variations for simulation times above \SI{100}{\s} were below \SI{10}{\%} of the measured times. \\
\\
CPU parallelisation using MPI was expected to reduce the simulation time linearly proportional to the number of cores with a factor just below unity (accounting for overhead). The GPU devices were expected to reduce the simulation time by several orders of magnitude and a preliminary study has demonstrated speed-up factors of upwards of one-thousand \cite{McXtrace3}.

\section{Resulting simulation times and speed-up factors}

This section considers the effect of the GPU acceleration for McXtrace instruments of increasing complexity. This section presents the measured run-time and speed-up factors for the three problems based on simulations of the three instruments: the basic instrument, the focusing instrument, and the EDXCT instrument.

\subsection{Parallelism for simple problems}

\begin{figure}[h!]
    \centering
    \begin{subfigure}[b]{0.49\textwidth}
        \centering
        \includegraphics[width=\textwidth]{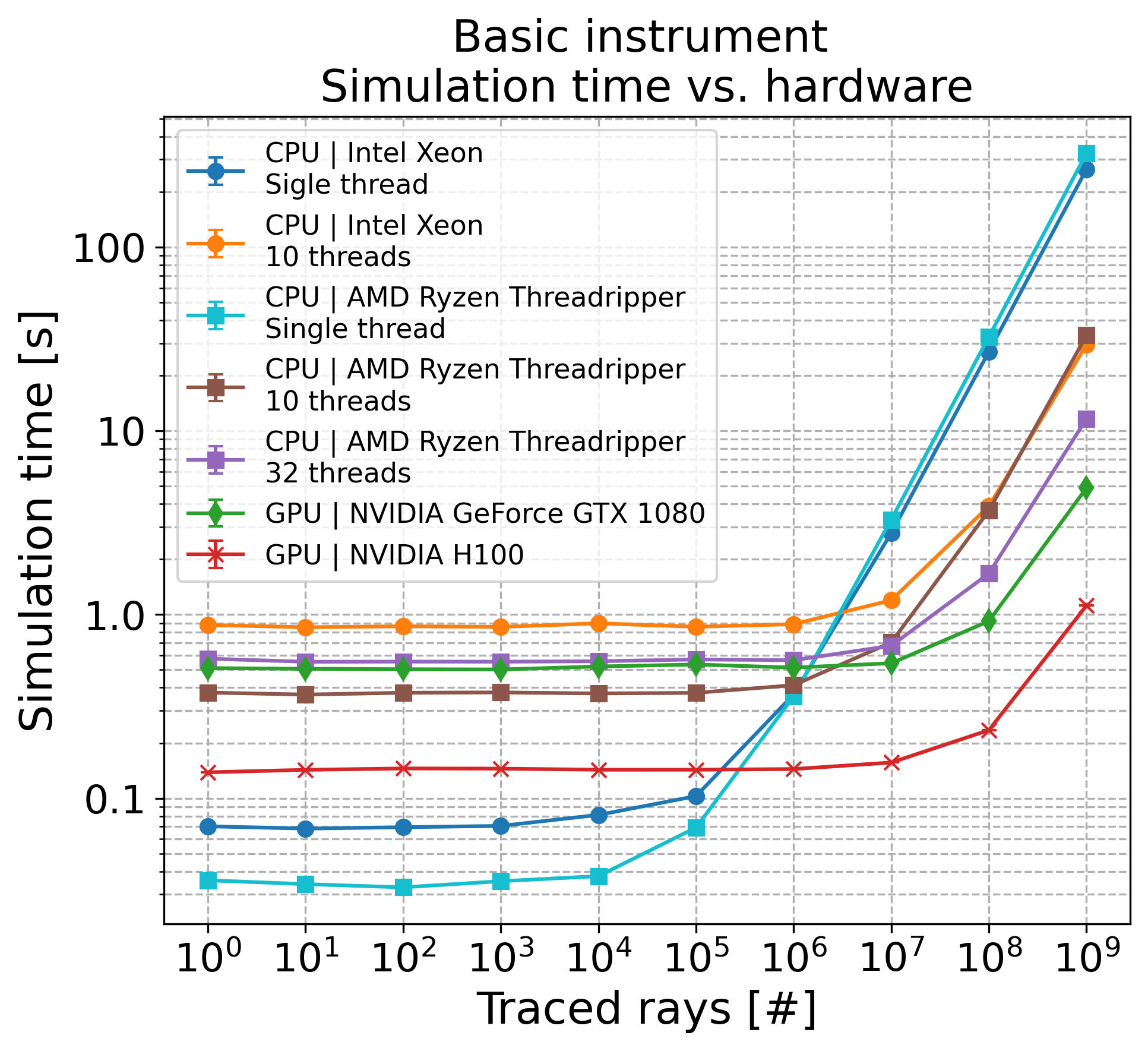}
        \caption{} \label{fig:4.1:a}
    \end{subfigure}
    \hfill
    \begin{subfigure}[b]{0.49\textwidth}
        \centering
        \includegraphics[width=\textwidth]{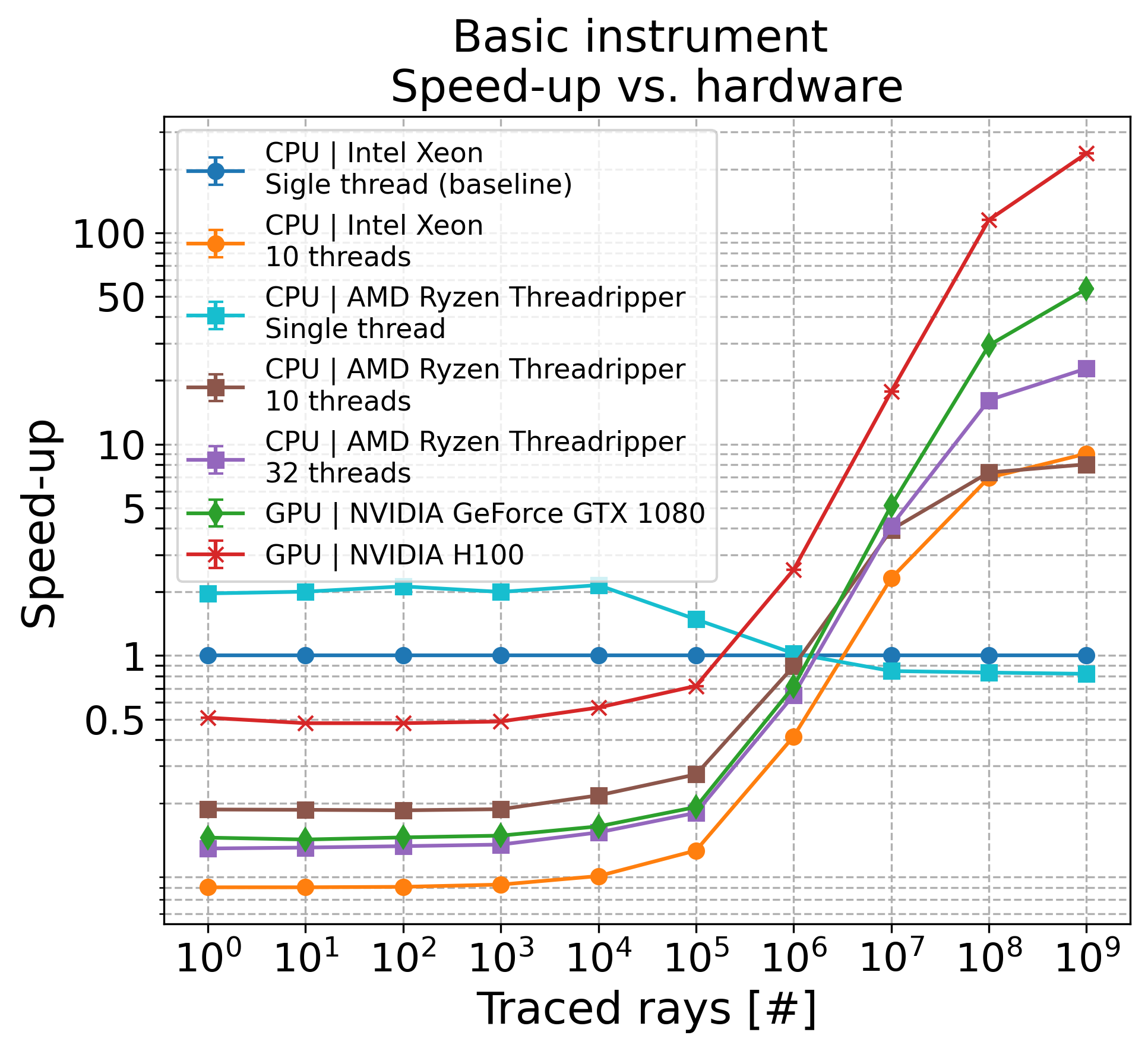}
        \caption{} \label{fig:4.1:b}
    \end{subfigure}
    \caption{Measured run-time (panel a) and relative speed-up factors (panel b) measured for simulations of the basic instrument in problem \ref{prob1}. Observations were done for varying problem sizes \SIrange{1}{e9}{} and measured for the hardware listed in table~\ref{tab:hardware}. The time measurements were based on the average of 100 repeated simulations at each level of count statistics. The time measurements of the single-threaded Intel Xeon CPU were used as the baseline for calculating the speed-up factors.} \label{fig:4.1}
\end{figure}

For the initial basic problem \ref{prob1}, Fig.~\ref{fig:4.1:a} reports on the measured run-time of simulating the basic instrument using the hardware specified in table~\ref{tab:hardware}. Fig.~\ref{fig:4.1:b} reports on relative speed-up normalised to the performance of a single CPU core. \\
\\
Figures \ref{fig:4.1:a}-\ref{fig:4.1:b} clearly show that up to a statistic of 1e7 rays, the effort of either CPU or GPU running in parallel does not yield major improvements to the simulation time. At higher statistics, the multi-threaded simulations run $\sim 10$ times faster than the single-threaded simulation (as expected) and the GPU simulation performs around \SIrange{50}{250}{} times faster than the single-threaded simulation dependent GPU model. 

\begin{figure}[h!]
    \centering
    \begin{subfigure}[b]{0.49\textwidth}
        \centering
        \includegraphics[width=\textwidth]{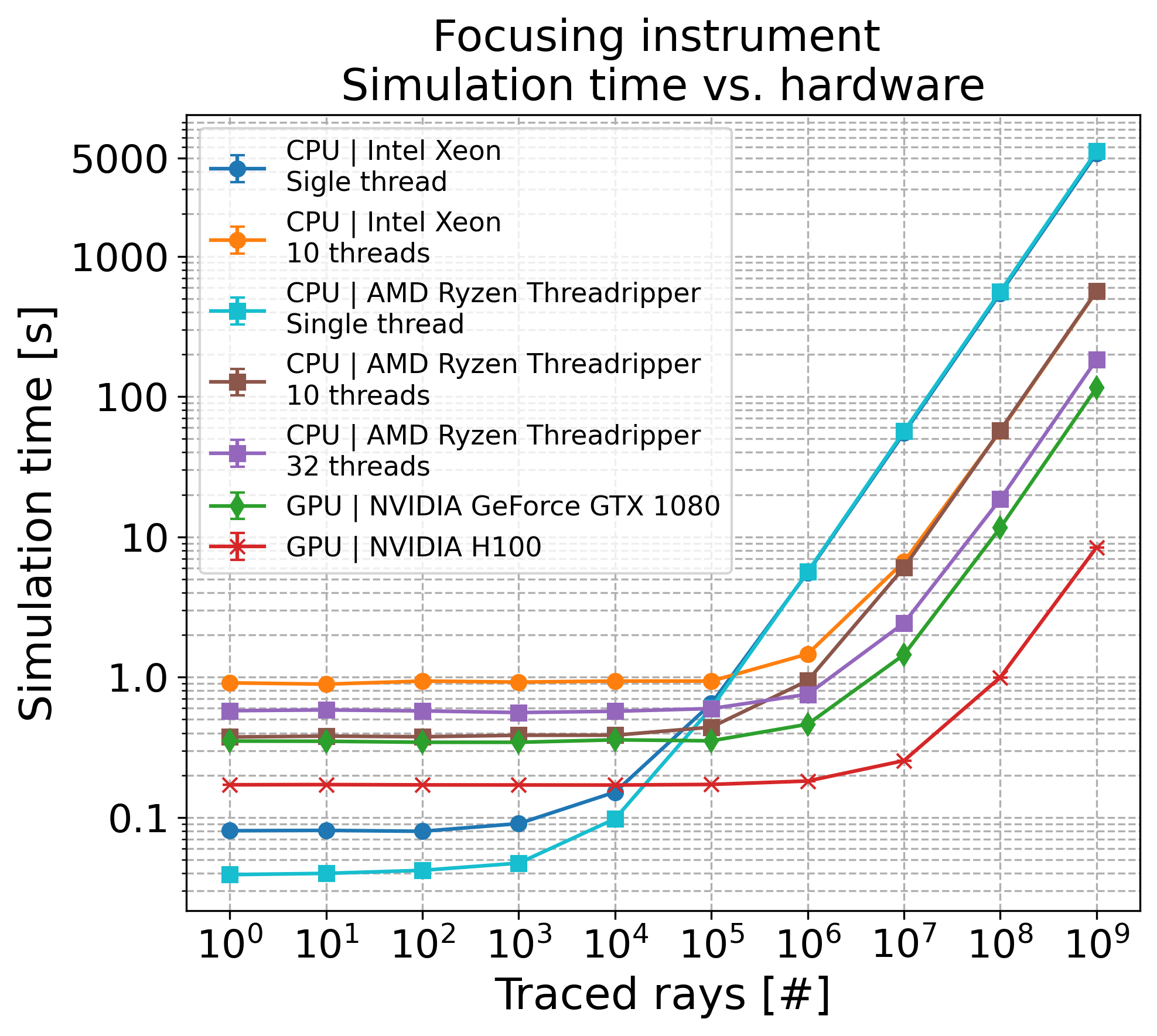}
        \caption{} \label{fig:4.2:a}
     \end{subfigure}
     \begin{subfigure}[b]{0.49\textwidth}
         \centering
         \includegraphics[width=\textwidth]{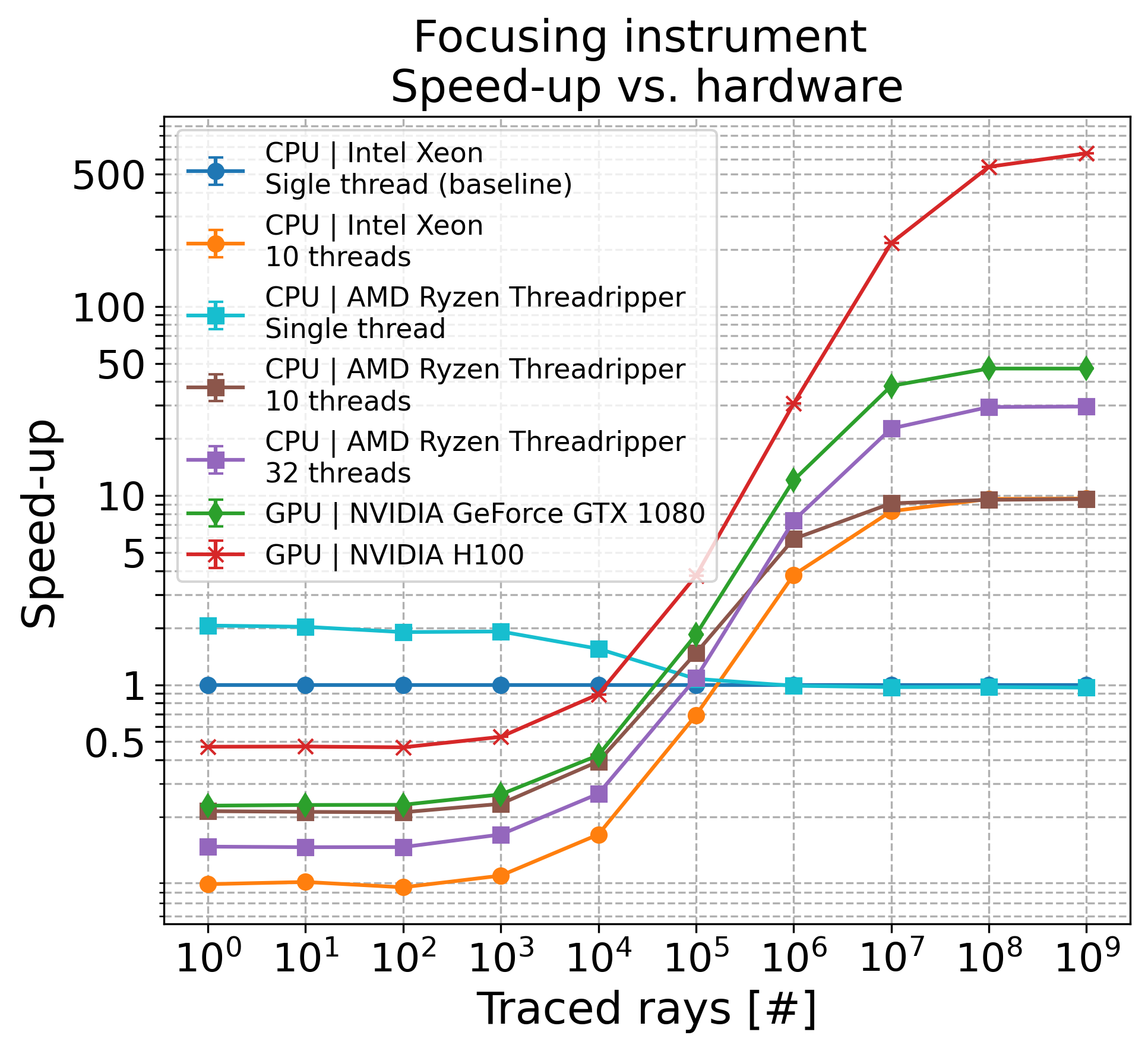} 
         \caption{} \label{fig:4.2:b}
     \end{subfigure}
     \caption{Measured run-time (panel a) and relative speed-up factors (panel b) measured for simulations of the focusing instrument in problem \ref{prob2}. Observations were done for varying problem sizes \SIrange{1}{e9}{} and measured for the hardware listed in table~\ref{tab:hardware}. The time measurements were based on the average of 100 repeated simulations at each level of count statistics. The time measurements of the single-threaded Intel Xeon CPU were used as the baseline for calculating the speed-up factors.} \label{fig:4.2}
\end{figure}

Problem \ref{prob2} introduces X-ray optics into the simulation, which increases the complexity and numerical demand of our simulation setup. The measured simulation time and speed-up factors for simulations of the focusing instrument are shown in figures \ref{fig:4.2:a} and \ref{fig:4.2:b}, respectively. The simulation times for the focusing instrument were around ten times longer than for the basic instrument. The parallelism begins to pay off slightly earlier than for the basic instrument at $\sim \SI{e5}{}$ rays. In agreement with problem \ref{prob1} we observe speed-ups corresponding to $\sim 10$ times for the MPI simulation and for GPU simulations upwards of \SIrange{50}{600}{} times depending on the GPU model.

\subsection{Accelerated ray-tracing of energy-dispersive computed tomography instruments}

\begin{figure}[h!]
     \centering
     \begin{subfigure}[b]{0.49\textwidth}
         \centering
         \includegraphics[width=\textwidth]{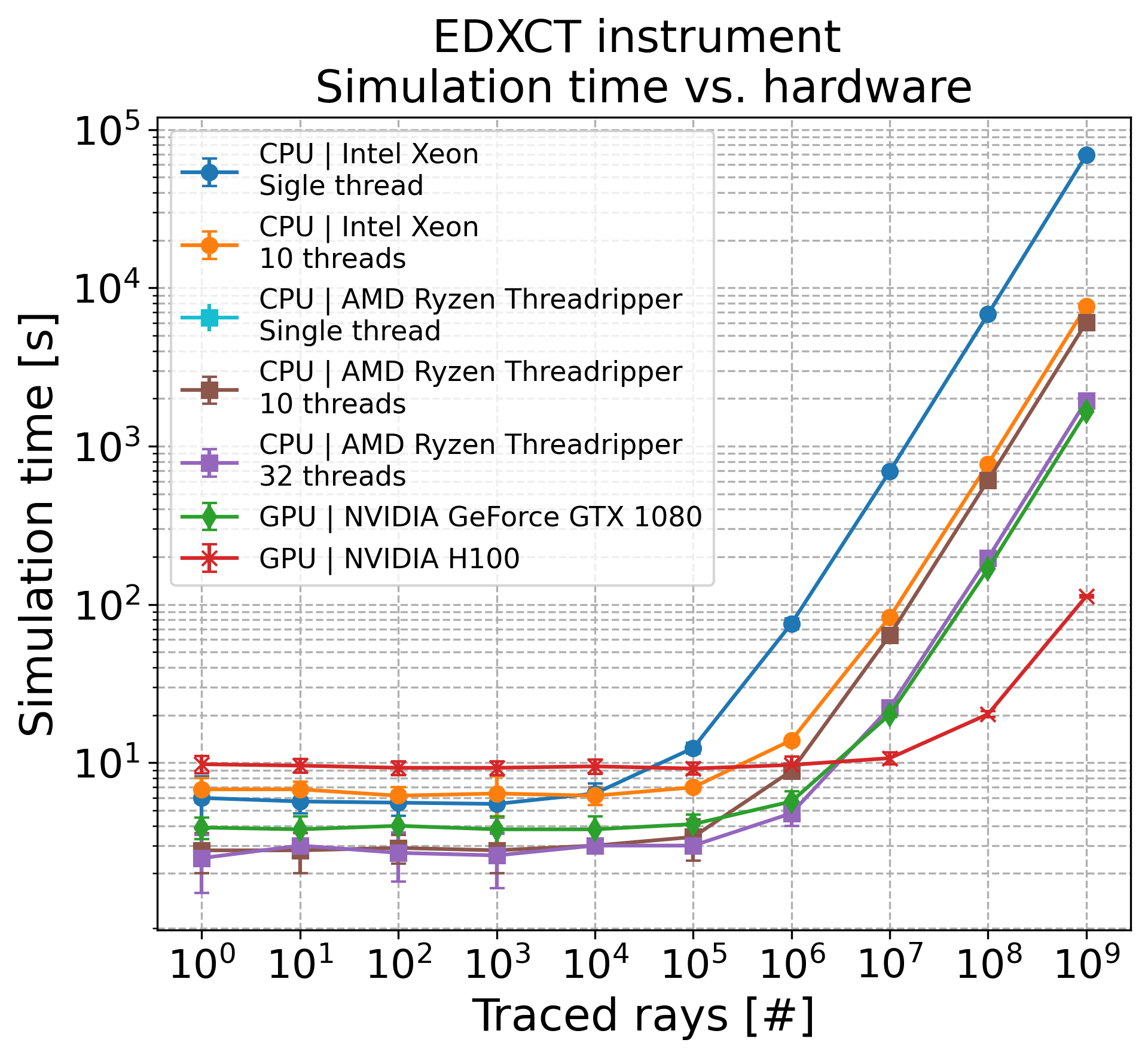}
         \caption{} \label{fig:4.3:a}
     \end{subfigure}
     \hfill
     \begin{subfigure}[b]{0.49\textwidth}
         \centering
         \includegraphics[width=\textwidth]{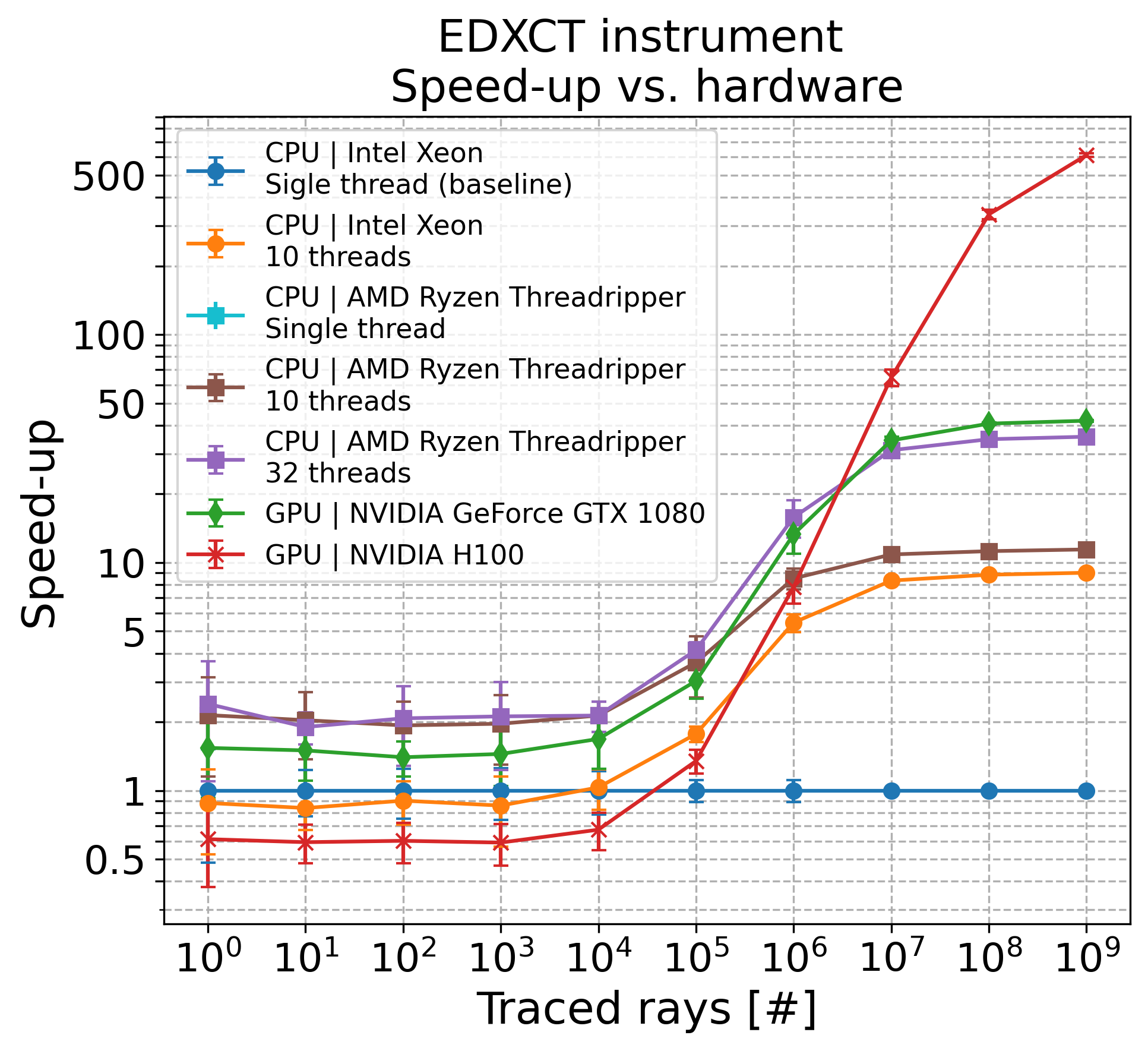}
         \caption{} \label{fig:4.3:b}
     \end{subfigure}
     \caption{Measured run-time (panel a) and relative speed-up factors (panel b) measured for simulations of the focusing instrument in problem \ref{prob2}. Observations were done for varying problem sizes \SIrange{1}{e9}{} and measured for the hardware listed in table~\ref{tab:hardware}. The simulations were repeated a maximum of 10 times, but some were simulated only once to keep the total simulation time below \SI{e5}{\s}. The variance of the time measurements did not exceed \SI{10}{\%} of the simulation time for statistics above \SI{e5}{}. The time measurements of the single-threaded Intel Xeon CPU were used as the baseline for calculating the speed-up factors.} \label{fig:4.3}
\end{figure}

The EDXCT instrument introduced for problem \ref{prob3} was more complex than both the basic and focusing instruments, and especially the \textit{Phantom\_sample3D} component increases the complexity by tracing the rays step-wise through the sample volume. Consequently, the simulation times shown in Fig.~\ref{fig:4.3:a} were around 100 and 10 times longer than for the basic and focusing instruments, respectively. The speed-up factors shown in Fig~\ref{fig:4.3:b} indicate that parallelisation begins to pay off for statistics above \SI{e5}{} rays, similarly to the focusing instrument. The GPU's speed-up factors seem to plateau at around \SIrange{40}{600}{} depending on the GPU hardware.  \\
\\
Preliminary investigations of the GPU-accelerated simulations of the EDXCT instrument revealed speed-up factors $\sim100$ times smaller than expected. An investigation of the GPU performance revealed inefficient memory handling, where a large amount of data ($\sim\SI{20}{TB}$) were transferred between the GPU and device memory during a simulation (input data size $\sim\SI{30}{MB}$). The memory access routine was modified to reduce the data transfers and obtain the results shown in figure~\ref{fig:4.3}.

\section{Discussion}
\label{sec:5}

This section starts with an estimation of the required count statistics and simulation time for performing energy-dispersive X-ray CT simulations for aviation security applications. This is followed by the discussion of some unexpected memory issues when running simulations on GPUs. Finally, a discussion of the resulting speed-up factors and how the results might be generalised to other McXtrace instruments.

\subsection{The required count statistics}

The number of rays ($n_{rays}$) required to probe all the spatial detector's pixels ($N_{pixel}$) increases linearly with the number of energy bins ($N_{bins}$). A lower bound can be estimated by $n_{rays}\leq N_{pixel} \cdot N_{bins}$, assuming a uniform distribution in space and energy of the detected rays. For the instrument in figure~\ref{fig:3:1:EDXCT_inst} with 5 detectors each with 128 pixels and 128 energy bins the number of rays becomes $n_{rays}\leq 5 \cdot 128 \cdot 128=81920 \lesssim 10^5$. \\
\\
For the EDXCT a better estimate of the number of rays can be achieved by considering the specific requirements of aviation security. Here we consider the aviation security quality control tool called the Standard Test Piece (STP) defined by the European Civil Aviation Conference (ECAC) in Doc 30 point 12.3.2 (Annex IV-12-M) \cite{ecac_stp}. Test number four of the STP tests the penetration thickness of a steel staircase with ten steps from \SI{14}{mm} up to \SI{30}{mm} (\SI{2}{mm} increments). A lead rod is placed beneath the centre of the staircase to provide a contrast for reference. The test result is accepted if the lead rod is visible beneath the \SI{26}{mm} steel step. \\
\\
The number of rays required to perform a simulation for aviation security applications can be estimated by the number of rays needed to identify the lead beneath \SI{26}{mm} steel. The STP is designed for visual inspection and does not provide mathematical models for evaluating the test results. We chose to consider the contrast-to-noise ratio (\textit{CNR}) between the mean intensity of regions of staircase material ($I_{mat}$) and reference material ($I_{ref}$), defined as:
\begin{equation}
    CNR = \frac{C}{N} = \frac{\left| I_{mat} - I_{ref} \right|}{\sqrt{\sigma_{mat}^2 + \sigma_{ref}^2}}
\end{equation}
where the contract (\textit{C}) and the noise (\textit{N}) were defined as the region's difference in mean intensity and summed squared standard deviations of the means ($({\sum_j\sigma_j^2})^{1/2}$), respectively. The simplest test criteria would be having a $CNR \geq 1$. However, due to stronger absorption at low X-ray energies, the test criteria were defined as having the linear fit of the \textit{CNR} (above some energy threshold) being greater than unity over the full energy range (\SIrange{20}{160}{\keV}). The flexibility of the instrument allows the staircase and reference materials to be chosen from a library of material properties and easily interchanged to test different materials. 

\begin{figure}[h!]
    \centering
    \includegraphics[width=0.75\linewidth]{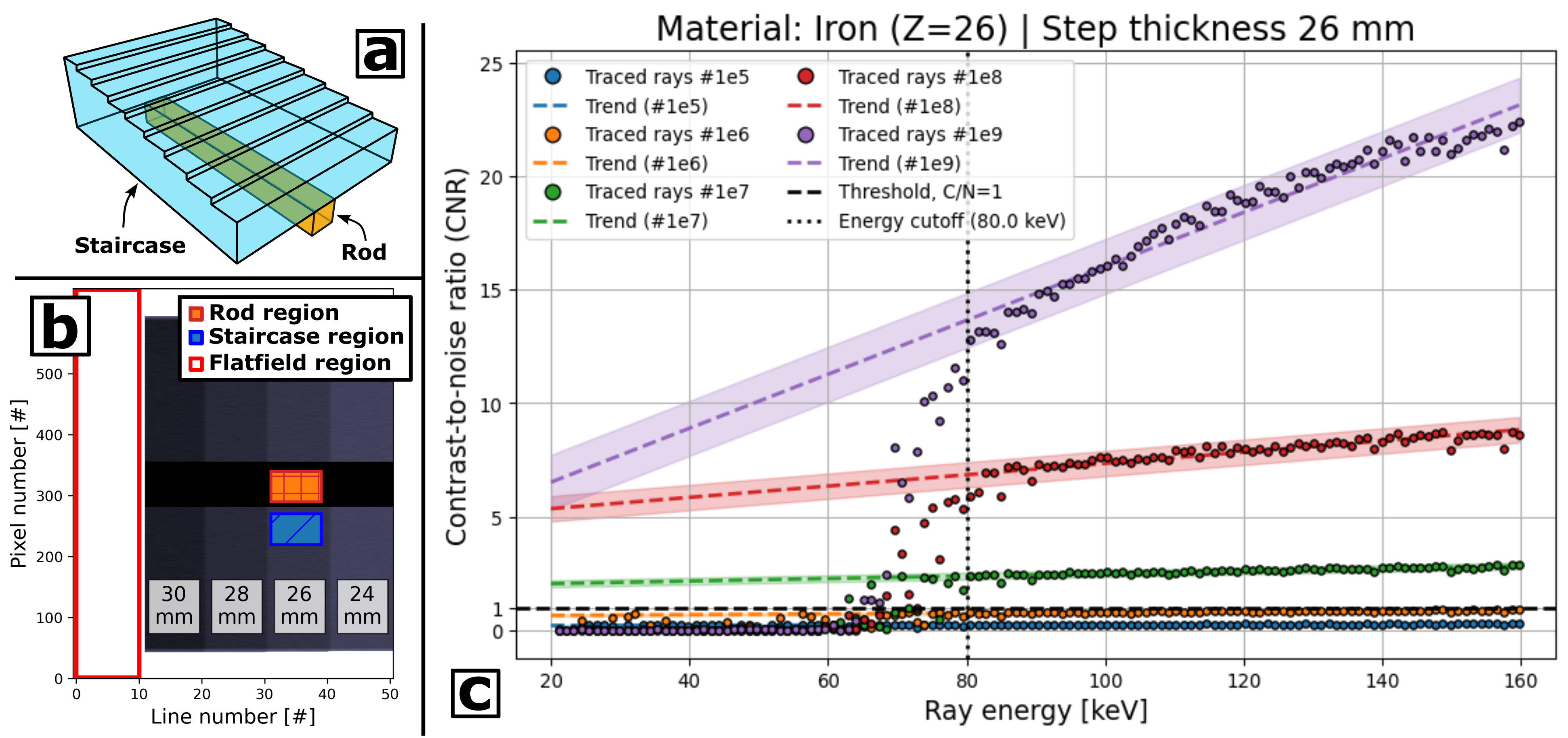}
    \caption{Panel a shows a 3D rendering of a staircase with a rod below. The staircase's steps range in height from \SIrange{14}{30}{mm}. Panel B shows a projected image of the four highest steps of a steel staircase simulated using the EDXCT instrument with \SI{e8}{} rays. The image intensity is normalised to the flatfield region. The staircase is oriented to generate a top-down view of the steps and the two coloured squares mark the regions used to estimate the contrast-to-noise ratio (CNR). The CNR for the two regions is plotted as a function of energy in panel C for simulations repeated at different levels of count statistics. }
    \label{fig:3.3.1:ana}
\end{figure}

The sample volume of a staircase shown in Fig.~\ref{fig:3.3.1:ana},a was input and simulated in the EDXCT instrument described in section \ref{sec:3.3} with the number of rays varying from \SI{e5}{} to \SI{e9}{}. The sample was configured with iron and lead as the staircase and reference materials. The two regions for the \SI{26}{\mm} thick step are sampled from the simulation (ten lines wide and ten pixels high), as shown in Fig.~\ref{fig:3.3.1:ana},b. The \textit{CNR's} are plotted in Fig.~\ref{fig:3.3.1:ana},c as a function of detected X-ray energy. A sharp change in the \textit{CNR} was observed at around \SI{80}{\keV}, due to strong absorption in the lead.  A least-squares-fit was used to fit the \textit{CNR's} for energies above the \SI{80}{\keV}. From $n_{rays}\geq \SI{e7}{}$ the \textit{CNR} was larger than unity over the energy range. \\
\\
Performing McXtrace simulations for aviation security applications using the EDXCT instrument would require $n_{rays}\geq \SI{e7}{}$ to ensure penetration of \SI{26}{mm} of steel. This estimation is here based on the evaluation test of a digital clone of the STP. This test does not include all the aspects of ensuring good image quality and it is particularly ill-suited for testing X-ray CT instruments. Yet, this analysis indicates the lower range of rays required to simulate a realistic use case. \\
\\
An estimate of the simulation time per line for a single-threaded CPU (Intel Xeon CPU) of \SI{e7}{} rays can be found in Fig.~\ref{fig:4.3} to be around $\SI{e3}{\s} \simeq \SI{15}{\min}$. The total simulation time per projection of a standard suitcase of \SI{55}{\cm} in length (with a \SI{1}{\mm} line spacing) become around $\SI{1500}{\hour}$. This is infeasible for testing various instrument configurations and geometries by repeated simulations. MPI parallelisation with 32-cores would reduce the time by around a factor of thirty to around $\SI{50}{\hour}$ per projection, which is still too long for simulating CT data with many projections. Only parallelisation using a high-performing GPU, like the NVIDIA H100, will reduce the time sufficiently to achieve a meaningful simulation time per projection below \SI{4}{\hour}. For CT applications, \SI{4}{\hour} per projection is still a long time when all full data sets contain hundreds of projections. However, for systems designed for a small number of projections ($<20$), like the setup modelled by the EDXCT instrument \cite{ss-euspen2023}, a full simulated data set can be acquired within a few days or a single day with 3-4 GPU cards in parallel. 

\subsection{General GPU-acceleration of McXtrace simulations}

McXtrace instrument with a short simulation time benefits the least from GPU acceleration as seen for the basic instrument with a peak speed-up factor of \SI{250}{}. For short simulations, the communication overhead reduces the speed-up factor as observed for the basic instrument needing more than \SI{e6}{} rays before parallelisation speed-up the simulations. Simple McXtrace instruments with low complexity gain the least from GPU implementation and parallelisation in general.\\
\\
The focusing and EDXCT instruments demonstrate peak speed-up factor upwards of \SI{600}{} for simulations run using the selected hardware. The EDXCT instrument's speed-up factor using the older NVIDIA GeForce GTX 1080 card was close to \SI{20}{\%} smaller than for the focusing instrument, almost tying with the AMD Threadripper (CPU) performance.  \\
\\
These results suggest instruments with a higher level of complexity, such as the focusing and EDXCT instruments, benefit the most from parallelisation. However, instruments with bottlenecks, e.g. frequent memory access, require more care to avoid memory issues, which could severely impact performance. Yet, carefully managing the code's memory access should produce results, similar to the EDXCT instrument simulated here. Speed-up factors on this scale are required for certain applications, such as for energy-dispersive CT simulations.

\section{Conclusion}
\label{sec:5}

Utilising parallel computing for simulating McXtrace instruments begins to reduce the simulation time for count statistics above \SIrange{e5}{e6}{} rays. Towards high statistics, the simulation time for the simple instrument scales linearly with the number of CPU cores as expected. The speed-up factors were around \SIrange{40}{50}{} for older (NVIDIA Geforce GTX 1080) and upwards of \SI{600}{} for modern (NVIDIA H100) GPU cards. \\
\\
The longer simulation time of the EDXCT instrument compared to the two simpler instruments is caused by the increased complexity and number of computations required for the step-wise propagation of the rays through the sample volume. The MPI speed-up still scales with the number of CPU cores and the GPUs are still major improvements with speed-up factors between \SIrange{40}{600}{}. The speed-up achieved using GPU acceleration is of great importance in some applications as demonstrated for energy-dispersive X-ray CT instruments within a practical time frame. \\
\\
These results are expected to be representative of other McXtrace instrument configurations with high complexity. For instruments requiring repeated memory access, the components' memory handling and access routines might need optimisation before achieving useful GPU speed-up factors.

\section*{Acknowledgements}
The authors would like to thank all developers from the joint McStas-McXtrace team for their contributions. A special acknowledgement goes to Erik B. Knudsen, Emmanuel Farhi, Mads Bertelsen and Jakob Garde who were all very active in GPU hackathons while the codes were adapted for OpenACC support. The authors would also like to thank Danilo Quagliotti and Leonardo De Chiffre for their contribution in discussing the research questions. \\
\\
This work was supported by the Innovation Fund Denmark (grant ref. number 1044-00087B) and the company Exruptive A/S.







\end{document}